\documentstyle[prl,aps,twocolumn,epsf]{revtex}
\def\break#1{\pagebreak \vspace*{#1}}
\begin{document}
\draft
\title{RKKY interaction and Kondo screening cloud
for strongly correlated electrons}
\author{Reinhold Egger$^1$ and Herbert Schoeller$^2$}
\address{${}^1$Fakult\"at f\"ur Physik, Albert-Ludwigs-Universit\"at,
Hermann-Herder-Stra{\ss}e 3, D-79104 Freiburg, Germany\\
${}^2$Institut f\"ur Theoretische Festk\"orperphysik,
Universit\"at Karlsruhe, D-76128 Karlsruhe, Germany  }
\maketitle
\widetext
\begin{abstract}
The RKKY law and the Kondo screening cloud
around a magnetic impurity are investigated for correlated 
electrons in 1D (Luttinger liquid). 
We find slow algebraic distance dependences,
with a crossover between both types of behavior.
Monte Carlo simulations have been developed to 
study this crossover. In the strong coupling regime, 
the Knight shift is shown to increase with
distance due to correlations.
\end{abstract}

\narrowtext
Since its discovery, the indirect Ruderman-Kittel-Kasuya-Yosida
(RKKY)  exchange interaction between localized magnetic impurities 
embedded in a host
metal has played an important role in the theory of magnetism.
The magnetic moment of one impurity scatters conduction
electrons, which are then seen by some other impurity.
This second-order process results in the $2k_F$-oscillatory
RKKY interaction between different magnetic moments\cite{rkky,kittel},
where $k_F$ is the Fermi momentum. For a lattice of magnetic impurities,
this interaction favors magnetically ordered phases
determined by the lattice geometry.
The RKKY interaction is a basic ingredient for
many phenomena in strongly correlated systems, 
e.g., magnetic impurities in quantum wires\cite{gogolin},
normal-state magnetism in high-temperature superconductors\cite{beck},
or magnetic ordering in heavy fermion materials 
\cite{brugger,loehneysen,fye87,affleck}. 

For uncorrelated conduction electrons, the RKKY interaction 
in $d$ dimensions is $\sim \cos[2k_F x]/x^d$,
where $x$ is the distance between the localized moments
\cite{rkky,kittel}. 
Within the  random-phase approximation or Fermi liquid theory,
this law is not expected to change qualitatively \cite{kittel}.
At the same time, however, it is known that Coulomb interactions
can modify spin- or charge-density correlation exponents 
in 1D \cite{luther,haldane,solyom,schulz,kane92,egger95}.
In this Letter, we show that the RKKY interaction indeed exhibits only
a slow algebraic decay  $\sim\cos[2k_F x]/x^{g_c}$, with an 
interaction-dependent exponent $g_c<1$. 
To examine how the RKKY law is affected
by the Kondo effect in such a strongly
correlated system, we study the magnetic
screening cloud around a single Kondo impurity.  We predict
the asymptotic behavior far away from the impurity
and qualitatively discuss how RKKY relates to
Kondo screening physics in a Luttinger liquid. 
The crossover between these two regimes has been 
analyzed by Monte Carlo (MC) simulation.

To describe the  low-energy properties of correlated 1D conduction 
electrons, we employ
the bosonization technique\cite{luther,haldane}. The 
spin-$\frac12$ electron field operator can equivalently be
expressed in terms of spin and charge boson fields,
which obey the algebra (we put $\hbar=1$)
\break{0.9in}
\begin{equation}
{[} \phi_i (x) ,\theta_j(x')] = 
-\frac{i}{2} \delta_{ij} \,{\rm sgn}(x-x') \;,
\end{equation}
where $i,j$ denote the charge ($c$) or spin ($s$) degrees of freedom.
The canonical momentum  for the $\theta_i$
phase field is  $\Pi_i(x)=\partial_x\phi_i(x)$.
Written in terms of the boson fields, the right- or left-moving component
($p=\pm$) of the electron annihilation operator for spin $\alpha=\pm$ 
is
\begin{eqnarray} \nonumber
\psi_{p\alpha} (x) &=& \frac{1}{\sqrt{2\pi a}}
 \,\eta_{p\alpha}
\exp\left[-i\sqrt{\pi/2}\left[\phi_c(x)+\alpha
\phi_s(x)\right]\right]
\\
\label{elcr} &\times& 
\exp\left[ ipk_F x+ ip\sqrt{\pi/2}
\left[ \theta_c(x)+\alpha \theta_s(x)\right]
\right] 
\;,
\end{eqnarray}
where $a=v_F/\omega_c$ is a short-distance cutoff
($\omega_c$ is the bandwidth cutoff, say, the
Fermi energy, and $v_F$ is the Fermi velocity).
The unitary zero-mode operators $\eta_{p\alpha}$ 
annihilate a particle from branch $p\alpha$ and ensure that
anticommutation relations hold between operators with
different $p\alpha$ \cite{haldane}. In contrast to models without 
spin flips, they have to be considered explicitly here to account 
for all minus signs.

The archetypical low-energy theory for correlated electrons in 1D
is the Luttinger liquid model \cite{haldane}, which unifies the
low-temperature physics of microscopic lattice models for 
strongly correlated fermions.
There are only two relevant interaction constants $g_c$ and $g_s$.
The charge interaction constant is $g_c\approx [1+2U/\pi v_F]^{-1/2} \leq 1$,
where $U$ is the forward-scattering amplitude of the screened Coulomb 
interaction potential. The Luttinger liquid  model assumes that 
one is away from half-filling so that Umklapp scattering is not present.
In addition, electron-electron backscattering processes 
are neglected, albeit one can incorporate them by a 
renormalization of the interaction constants or by a perturbative
renormalization group (RG) scheme, where the fixed-point value is given by
$g_s=1$ \cite{schulz}. Therefore, we will put
$g_s=1$ in the following to respect the underlying SU(2) spin symmetry
of the electrons.  The Hamiltonian of the clean system is then given by 
\begin{equation}
H_0= \sum_{j=c,s} \frac{v_j}{2} 
\int dx \left[ g_j \Pi_j^2 + g_j^{-1} ( \partial_x
\theta_j )^2\right]\;,
\end{equation}
where $v_j=v_F/g_j$ is the velocity of charge or spin density waves
for the case of full Galilean translational invariance considered
here \cite{schulz}.

Let us now add a spin-$\frac12$ magnetic impurity at $x=0$.
We use the standard contact term with the conduction electrons,
$H_I=J \vec{s}(0)\vec{S}$ \cite{kondoll,emery},
where $J$ is the direct exchange coupling,
$\vec{S}$ the impurity spin operator, and $\vec{s}(x)$ the
spin density operator, which from Eq.~(\ref{elcr}) reads in 
bosonized form 
\begin{eqnarray} \nonumber
s_z(x)& =&  \frac{\partial_x \theta_s}{\sqrt{2\pi}} 
+\frac{\sigma_z}{\pi a} \cos[2k_Fx + \sqrt{2\pi}\,\theta_c(x)]
\cos[\sqrt{2\pi}\,\theta_s(x)] \\
s_\pm(x) &=& \frac{1}{\pi a} \exp[ \pm\sqrt{2\pi}\,i\phi_s (x) ]
\Bigl\{\pm i\sigma_y \cos[\sqrt{2\pi}\,\theta_s(x)] + \nonumber \\
&+& \sigma_x \cos[2k_F 
x+\sqrt{2\pi}\,\theta_c(x)]\Bigr\}\;,
\end{eqnarray}
with $s_\pm = s_x \pm i s_y$.
Here we have used that the $\eta_{p\alpha}$
show up only as bilinear
forms, for which a convenient representation can 
be found in terms of Pauli matrices,
\begin{equation} 
\eta^\dagger_{p,\alpha} \eta^{}_{-p,\alpha} \rightarrow
 \alpha \sigma_z \;, \;
\eta^\dagger_{p,\alpha} \eta^{}_{p,-\alpha} \rightarrow
 i\alpha\sigma_y\;, \;
\eta^\dagger_{p,\alpha} \eta^{}_{-p,-\alpha} \rightarrow
 \sigma_x \;.
\end{equation}
This replacement gives the correct sign for all possible
products of $\eta^\dagger_{p\alpha}\eta^{}_{p^\prime\alpha^\prime}$ pairs
allowing for a nonvanishing contribution. Therefore
the chosen representation is sufficient for the calculation of
correlation functions involving only spin or charge 
densities \cite{majorana}.

To eliminate an explicit dependence of the interaction part $H_I$
on the $\phi_s$ field, we
perform a standard unitary transformation \cite{emery},
$U= \exp[\sqrt{2\pi} i \phi_s(0) S_z]$,
such that our final Hamiltonian reads
\begin{eqnarray} \nonumber
&& UHU^{-1} = H_0 + \frac{\bar{J}}{\sqrt{2\pi}} S_z 
\partial_x \theta_s(0)
+\frac{J}{\pi a} 
 \Bigl( \sigma_x S_x \cos[\sqrt{2\pi}\theta_c] 
\\ &&+
\; \label{hamilt}
\sigma_y S_y \cos[\sqrt{2\pi}\theta_s]
+ \sigma_z S_z \cos[\sqrt{2\pi} 
\theta_s] \cos[\sqrt{2\pi} \theta_c] \Bigr)_{x=0} \;,
\end{eqnarray}
where $\bar{J} = J - 2\pi v_F$. 
The four interaction terms include
two forward and two backward scattering terms with or without spin
flip, respectively. Backward scattering
$(\sim \cos[\sqrt{2\pi}\theta_c])$ is responsible for RKKY oscillations,
while Kondo screening arises due to spin flip
terms $(\sim S_{x/y})$.  

Our subsequent discussion is based on
the correlation function $C(x)= \langle s_z(x) S_z\rangle$,
where the brackets indicate a thermal average. 
This function describes the spatial 
correlation of the electron spin density
with the impurity spin.
Another impurity spin located at $x$ would see this correlation,
and lowest-order perturbation theory in $J$ constitutes an 
exact derivation of the RKKY law\cite{kittel}.  
While a quantitative discussion of the complicated interplay between
the RKKY interaction and Kondo screening behaviors requires a study
of higher-order terms in the corresponding two-impurity model 
\cite{fye87,affleck}, the main qualitative features
of this interplay can already be extracted from $C(x)$ \cite{fye87}.
In this work, we therefore focus on the local screening properties induced
in a Luttinger liquid by the presence of a single impurity. 
A related quantity of direct experimental relevance
is the local susceptibility $\chi(x)=\partial\langle s_z(x)\rangle
/\partial B$, which was recently reconsidered
for the uncorrelated case \cite{sorensen} and is proportional to 
the Knight shift. 
Linear response theory gives ($\beta=1/k_B T$)
\begin{equation}\label{knight}
\chi(x) = \beta C(x) +  \beta \int dx^\prime \langle s_z(x) s_z(x^\prime)
\rangle \;,
\end{equation} 
where the second part does not contribute in the perturbative RKKY regime. 

Since the slowly varying part of $C(x)$ leads only to subleading terms
for $x\gg a$ \cite{sorensen,foot11}, 
we restrict ourselves to the $2k_F$ part in the following.
After the unitary transformation, we obtain 
\begin{equation} \label{cx}
C(x)  =  \frac{\cos[2k_F x]}{2\pi a} 
\langle \sigma_z \sin[\sqrt{2\pi} \theta_s(x)] 
\cos[\sqrt{2\pi} \theta_c(x)]\rangle\;.
\end{equation}
As can be seen from Eq.~(\ref{cx}), there is no phase shift
in the $\cos[2k_F x]$ term. This is clear
since Eq.~(\ref{hamilt}) does not include
elastic potential scattering  by the impurity.

The standard treatment of the RKKY interaction \cite{kittel} 
corresponds to a calculation of the correlation function
$C(x)$ by lowest-order perturbation theory in the exchange
coupling $J$. 
The finite-temperature result is ($x\gg a)$
\begin{eqnarray}\label{pert}
C(x) &=&  - \frac{1}{8a}\frac{J}{2\pi v_F} 
\left(\frac{\beta \omega_c}{\pi}\right)^{-g_c} \cos[2k_F x] 
\\ &\times & \nonumber 
 \int_0^{\beta} \frac{d\tau}{\beta}
\prod_{j=c,s}
\left| \sin\left[ \frac{\pi}{\beta}(\tau+ix/v_j) \right] \right|^{-g_j}
 \;.
\end{eqnarray}
For $x \ll x_T$, where $x_T= v_F/k_B T$ denotes the 
thermal lengthscale, this yields the RKKY law
\begin{equation} \label{rkky}
C(x) \sim - \frac{1}{a} \frac{J}{2\pi v_F} 
\cos[2 k_F x]  (x/a)^{-g_c} \;,
\end{equation}
while for  $x \gg x_T$, an exponential decay on the  
scale $x_T$ is obtained.  It is obvious that spin-flip
events do not contribute to the perturbative result (\ref{pert}).
Therefore,  to lowest order in $J$, one could just as well
consider a static impurity,
or, equivalently, a point-like magnetic field acting at $x=0$.
The presence of such a field induces $2k_F$-periodic
oscillations in the spin density of the electrons, which
are then responsible for the RKKY interaction. 
Thus the range function \cite{kittel}
describing the decay of the RKKY oscillation amplitude
displays only a slow {\em algebraic} $\sim
x^{-g_c}$ law in the low-temperature regime $x\ll x_T$.
In the noninteracting case, $g_c=1$, the usual $x^{-1}$
decay is recovered.
This modification of the range function might come as a
surprise, since the Coulomb interaction does not couple to 
spin densities. The slower decay is a many-body effect
induced by the presence of correlations.

Starting from order $J^2$ on, spin flips contribute 
and it becomes mandatory to treat the dynamics
of the impurity spin.  The most important aspect of the impurity
dynamics is the Kondo effect, leading to a screening
of the impurity spin by the Luttinger liquid spin density below
the Kondo temperature \cite{kondoll},
$T_K \sim J^{2/(1-g_c)}$.
Kondo screening of the impurity becomes important 
for strong couplings $J$ or at low temperatures.
For instance, the second-order contribution 
to $C(x)$ at $x\ll x_T$ is
\begin{equation}
\delta C(x)\sim - \frac{1}{2\pi a} \left(\frac{J}{2\pi v_F}\right)^2
\cos[2k_F x] (x/a)^{-g_c} \ln(x/a) \;.
\end{equation}
The logarithmic corrections over Eq.~(\ref{rkky}) are typical
for the Kondo effect and indicate that we are dealing 
with a nonperturbative problem.

To study the crossover from the RKKY law 
to the Kondo screening cloud, we have developed MC simulations. 
Since the nonlinear terms in Eq.~(\ref{hamilt})
are local, we integrate out all fields away from $x=0$. Under a path-integral
representation, we can rewrite $C(x)$ as an average over new fields $q_j(\tau)
=\sqrt{2\pi} \theta_j(0,\tau)$,
where $j=c,s$ and $\tau$ is the Euclidean time, and over the
impurity spin field $S(\tau)= 2 S_z(\tau) = \pm 1$. The
$\sigma_{x,y,z}$ operators
have to be treated dynamically as well, but from Eq.~(\ref{hamilt})
it follows that the corresponding field is constrained to be
$\sigma_z(\tau)=\mu S(\tau)$ with $\mu=\pm 1$.
We find the formal result
\begin{equation} \label{cs2}
C(x) = -\frac{1}{2\pi a}  \cos[2k_F x] \, W_c(x) W_s(x)\, D(x) \;.
\end{equation}
The functions $W_j(x)$
describe an algebraic decay $\sim x^{-g_j/2}$ on scales
$x\ll x_T$, followed by a crossover to an exponential decay,
\begin{equation} \label{wx} 
W_j (x) = \left( \frac{\beta \omega_c}{\pi} \sinh
\left[ \frac{2\pi x}{\beta v_j}  \right] \right)^{-g_j/2} \;.
\end{equation}
The impurity average is now contained in
\begin{eqnarray} \label{dx}
D(x) &=& - \Biggl\langle \mu S(\tau=0) 
\cos\left[ \frac{1}{\beta} \sum_\omega e^{-|\omega x|/v_c} q_c(\omega)
\right ] \\ &\times& \nonumber
\sin\left[ \frac{1}{\beta}
\sum_\omega e^{-|\omega x|/v_F} \left( q_s(\omega)- \frac{\bar{J}}
{4 v_F} S(\omega) \right)\right ]
 \Biggr\rangle \;,
\end{eqnarray}
where the average is taken using the action
\begin{eqnarray} \label{sss}
S&=& \sum_{j=c,s} \sum_\omega
 \frac{|\omega|}{2\pi g_j \beta} |q_j(\omega)|^2
+ S_J \\ &+&\frac{\pi }{2\beta}     \nonumber
 (\bar{J}/4\pi v_F)^2  \sum_\omega |\omega| 
|S(\omega)|^2 \;.
\end{eqnarray}
Frequency sums run over Matsubara frequencies,
and $q_j(\omega)$ and $S(\omega)$ are the Matsubara components of
the respective fields. 
Discretizing Euclidean time into $N$ slices, $\tau_j = j\Delta \tau$
with $\Delta \tau=\beta/N$, the part $S_J$ becomes
\begin{equation}
e^{-S_J} = \lim_{N\to \infty}\prod_{i=1}^N \langle \mu
S_{i+1},S_{i+1}| \exp[-\Delta\tau  H_J(\tau_j)]
| \mu S_i,S_i\rangle \;,
\end{equation}
where $H_J(\tau)$ is the last part ($\sim J$) of the Hamiltonian 
(\ref{hamilt}), with $\sqrt{2\pi}\theta_{c/s}(0)$ being replaced by
$q_{c/s}(\tau)$.
The matrix elements can be evaluated in closed form,
with $\sigma_z$ parametrized by $\mu S_i$ with 
$S_i=S(\tau_i)=\pm 1$.  
Since $\exp[-S_J]$ is negative for
certain impurity spin paths, 
our simulation method has to deal with the conventional
sign problem\cite{loh}. Fortunately, the sign problem is moderate
except near $T=0$. 

\begin{figure}
\epsfysize=6.5cm
\epsffile{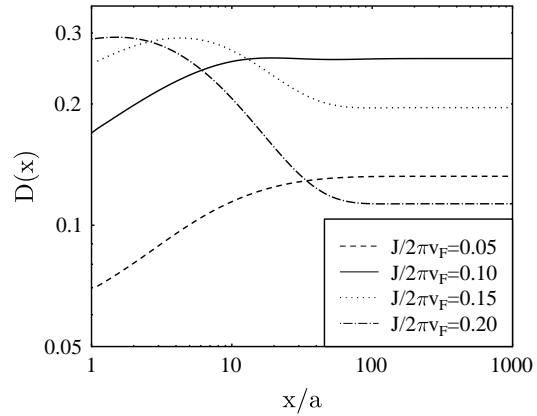}
\caption[]{\label{fig1}
 Monte Carlo data for $D(x)$ at $g_c=1/2$
and $\beta\omega_c=100$. Statistical errors are of the
order $5 \%$. Notice the logarithmic scales.}
\end{figure}
\begin{figure}
\epsfysize=6.5cm
\epsffile{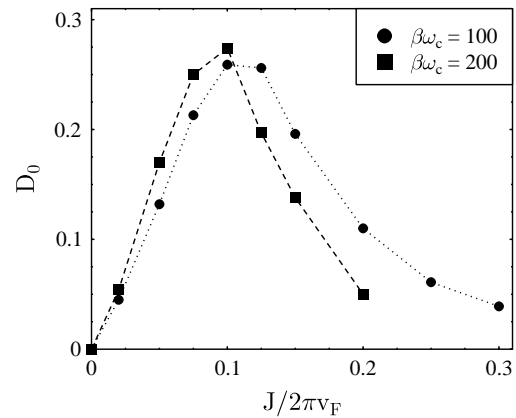}
\caption[]{\label{fig2}
Numerical results for the plateau value $D_0$ as
a function of $J$ for $g_c=1/2$ and two different temperatures.
Statistical errors are of the order of the symbol sizes, and
dotted and dashed lines are guides to the eye only.}
\end{figure}

In Fig.~\ref{fig1}, MC data for $D(x)$ 
are shown for several $J$ at $g_c=1/2$.
For small $J$, the function $D(x)$ exhibits a power law $x^\delta$
for $x\ll x_T$, where $\delta$ coincides with the RKKY law.
Far away from the impurity, $D(x)$ reaches a plateau value $D_0$ in
agreement with Eq.~(\ref{pert}). Therefore the RKKY law
 is fully reproduced by our simulations.
For large $J$, the numerical results display a different behavior.
The function $D(x)$ now  decreases to a small
plateau value, and the RKKY law breaks down even at short
lengthscales.  From our numerical data, one has a complete breakdown 
of RKKY for $J>J^*$ with $J^*/2\pi v_F\approx 0.1$. Furthermore,
the numerical simulations predict the asymptotic exponent $(1+g_c)/2$,
since $D(x)$ generally reaches its plateau value $D_0$ for $x < 
x_T$. From Eqs.~(\ref{cs2}) and (\ref{wx}),
one then infers the asymptotic form of $C(x)$, 
\begin{equation} \label{sc}
C(x)\sim \cos[2k_F x] (x/a)^{-(1+g_c)/2}\;,\;v_F/T_K\ll x\ll x_T\;,
\end{equation}
which we have also verified by using lower
simulation temperatures than in Fig.~\ref{fig1}.

In view of Fig.~\ref{fig1}, it seems convenient to discuss
the suppression of RKKY oscillations by Kondo screening
in terms of $D_0$.  Numerical results for the plateau value $D_0$
at $g_c=1/2$ are shown in Fig.~\ref{fig2}.
Taking some fixed $J<J^*$ and then going to low temperatures 
leads to  an increase in $D_0$. On the other hand, for $J>J^*$,
we observe a decrease in $D_0$ with lower temperatures. 
This indicates a crossover from a regime $J<J^*$, where RKKY
behavior is observed, to a non-RKKY regime $J>J^*$. 
Finally, for the special value $J=2\pi v_F$ (Toulouse limit),
i.e., $\bar{J}=0$,
one finds the exact result $C(x)=0$ implying that $D_0\to 0$
as $J$ approaches the Toulouse limit. The correlation
function $C(x)$ vanishes identically since the Hamiltonian (\ref{hamilt})
stays invariant under the transformation $\theta_s(x)\to
- \theta_s(x)$, whereas Eq.~(\ref{cx}) changes sign.

It is instructive to compare 
 the asymptotic behavior (\ref{sc}) of $C(x)$ with the Friedel oscillation
of the charge density.
Renormalization group and conformal field theory 
imply that in the strong-coupling limit, $S_z$ and $s_z(0)$ form
a local singlet \cite{kondoll}.
 This singlet decouples from the system and simply
acts as an elastic potential scatterer in the unitary limit. The Friedel 
oscillation for that case is given in Ref.\cite{egger95}. 
In a magnetic field $B$ one obtains for spin $\sigma=\pm$
\begin{equation} 
\label{fri}
\rho_\sigma(x) = \frac{k_F^\sigma}{\pi} - \frac{\sin[2k_F^\sigma x]}
{2\pi \alpha_\sigma} (x/\alpha_\sigma)^{-(1+g_c)/2}\;,
\end{equation}
where $k_F^\sigma = k_F + \sigma B/4v_F$
and $\alpha_\sigma=1/2g_c k_F^\sigma$. Clearly,
the Friedel oscillation and $C(x)$ are both characterized by the 
same asymptotic exponent.

As demonstrated in Ref.\cite{sorensen},
the Friedel oscillation can also be employed
to determine the $T=0$ local susceptibility (\ref{knight}). This 
quantity is experimentally accessible in terms of the Knight shift.
Using $\chi(x)=\partial\langle s_z(x) \rangle /\partial B$ and
$\langle s_z(x) \rangle = \sum_\sigma \sigma \rho_\sigma (x)/2$,
we obtain the leading asymptotic behavior,
\begin{eqnarray}\label{knight2}
4\pi v_F \chi(x) &= & -(x/a_0)^{(1-g_c)/2} \cos[2k_F x]  + 1
\\ &-& \frac{g_c(1-g_c)}{2}
\sin(2k_F x) (x/a_0)^{-(1+g_c)/2} \;,\nonumber
\end{eqnarray}
where $a_0=1/2g_c k_F$.
Remarkably, for correlated conduction electrons, the 
Knight shift actually {\em increases}
with distance. A related behavior has been reported for 
a non-magnetic impurity in a Heisenberg chain \cite{eggert}.

To conclude, for correlated electrons,
the RKKY interaction exhibits only a slow algebraic decay. 
This implies that the usual logarithmic $2k_F$-singularity of
the 1D susceptibility is turned into an algebraic divergence.
Furthermore, there is an interesting crossover from
RKKY to Kondo screening cloud behavior. 
Both are characterized by different exponents, and
both lead to a slower decay than in the noninteracting case.

We wish to thank J. von Delft, H. Grabert, and P. W\"olfle 
for useful discussions. H.S.~acknowledges support by the
Deutsche Forschungsgemeinschaft (SFB 195).


\begin{references}
\bibitem{majorana} For single-particle correlation functions, 
one would have to choose a representation for the
$\eta_{p\alpha}$ themselves.
\bibitem{foot11} 
The local formation of a Kondo singlet leads 
to a $\delta(x)$-contribution to $C(x)$ which is not specified here.
\end{references}
\end{document}